\documentclass[12pt]{article}
\usepackage{amsfonts}
\usepackage{eurosym}
\usepackage{graphicx}
\usepackage[utf8]{inputenc}
\usepackage[T1]{fontenc}
\usepackage{indentfirst}
\usepackage[margin=0.6in,nomarginpar]{geometry}
\usepackage[final]{hyperref}
\usepackage{amsmath}
\usepackage{hyperref}
\usepackage{cite}
\usepackage{subcaption}
\usepackage{caption}
\usepackage{amssymb}
\usepackage{multirow}
\usepackage[table]{xcolor}
\usepackage{orcidlink}
\usepackage{authblk}
\usepackage{float}
\hypersetup{
colorlinks=true,
linkcolor=blue,
citecolor=blue,
filecolor=magenta,
urlcolor=blue
}

\begin{document}

\title{Dunkl-Schr\"odinger Equation in Higher Dimensions}
\author{B. Hamil \orcidlink{0000-0002-7043-6104} \thanks{%
hamilbilel@gmail.com/bilel.hamil@umc.edu.dz (Corresponding author)} \\
Laboratoire de Physique Mathématique et Subatomique, \\
Faculté des Sciences Exactes, Université Constantine 1 Frères Mentouri, Constantine, Algeria.\\
 \and B. C. L\"{u}tf\"{u}o\u{g}lu %
\orcidlink{0000-0001-6467-5005} \thanks{%
bekir.lutfuoglu@uhk.cz } \\
Department of Physics, Faculty of Science, University of Hradec Kralove, \\
Rokitanskeho 62/26, Hradec Kralove, 500 03, Czech Republic. \and M. Merad 
\orcidlink{0000-0001-7547-6933} \thanks{
meradm@gmail.com} \\
Laboratoire de systèmes dynamiques et contr\^{o}le (L.S.D.C), \\
Département des sciences de la matière, Faculté des Sciences Exactes\\et SNV, Université de Oum-El-Bouaghi, 04000, Oum El Bouaghi, Algeria.}
\date{}
\maketitle

\begin{abstract}
This paper presents analytical solutions for eigenvalues and eigenfunctions of the Schrödinger equation in higher dimensions, incorporating the Dunkl operator. Two fundamental quantum mechanical problems are examined in their exact forms: the d-dimensional harmonic oscillator and the Coulomb potential. In order to obtain analytical solutions to these problems, both Cartesian and polar coordinate systems were employed. Firstly, the Dunkl-Schrödinger equation is derived in d-dimensional Cartesian coordinates,  and then for the isotropic harmonic potential interaction, its solutions are given. Subsequently, using polar coordinates the angular and radial parts of the Dunkl-Schrödinger equation are obtained. It is demonstrated that the system permits the separation of variables in both coordinate systems, with the resulting separated solutions expressed through Laguerre and Jacobi polynomials. Then, the radial Dunkl-Schrödinger equation is solved using the isotropic harmonic, pseudoharmonic, and Coulomb potentials. The eigenstates and eigenvalues are obtained for each case and the behavior of the energy eigenvalue functions are illustrated graphically with the reduced probability densities.

\end{abstract}

\section{Introduction}
Quantum mechanics is a branch of mathematical physics that deals with the behavior of very small-scale atomic and subatomic systems, which cannot be described successfully within the framework of classical mechanics. The investigation of such systems necessitates the utilization of differential equations that are tailored to the characteristics of the particles under consideration. One such equation is the Schrödinger equation, which is frequently employed to examine the dynamics of non-relativistic systems irrespective of the spin feature of the system. The Klein-Gordon and Dirac equations are commonly deployed to investigate relativistic systems with spin-$0$ and $1/2$ features, respectively. In all of these fundamental equations, the conventional Heisenberg algebra or the Dirac algebra is satisfied. 

It is worthy of note that the Dunkl formalization, which is integrated into the aforementioned equations and classifies the solutions according to the reflection symmetry, has been the subject of increasing interest, particularly over the past decade. The Dunkl formalization is founded upon the utilization of the Dunkl operator, which represents an extension of the ordinary partial spatial derivative with the incorporation of an additional term that involves the differential-difference operators associated with the reflection group \cite{Dunkl1989}. However, it should be noted that this extension results in a deformation of the usual algebra. Although the Dunkl operator, named after the pure mathematician Charles Dunkl, who proposed it in 1989, has surprisingly its origins in the middle of the last century. The physicist Wigner, rather than deriving equations of motion from commutation relations, discussed whether commutation relations could be derived from equations of motion with the opposite idea \cite{Wigner1950}. To gain insight into this question, Wigner examined the classical harmonic oscillator and the free particle problems and concluded that an inverse relationship could not be always unique. The following year, Lee investigated the same question for the quantum harmonic oscillator and demonstrated that Wigner's proposal could always be realized by utilizing a reflection operator and a deformation parameter in a well-defined Hilbert vector space \cite{Yang1951}. Subsequently, Green utilized Yang's deformed algebra to propose a generalized field quantization \cite{Green1953}. This approach proved to be of significant benefit in facilitating the introduction of color degrees of freedom and quantum chromodynamics \cite{Greenberg1964}. 

Despite the pervasive deployment of the Dunkl operator in mathematical research, its utilization in the field of physics was largely circumscribed to specific domains until the last decade \cite{Chakra1994, Lapointe1996, Kakei1996, Plyushchay1994, Plyushchay1996, Plyushchay1997, Gamboa1999, Plyushchay2000, Klishevich2001, Hovarthy2004, Rodrigues2009, Horvathy2010, Bie2012, Correa2014}. By 2013, this trend underwent a significant transformation. Initially, Genest et al. conducted an in-depth examination of the isotropic, anisotropic, and supersymmetric Dunkl oscillators in two- and three-dimensional space \cite{Genest20131, Genest20132, Genest20133, Genest20141, Genest20142}. These papers were succeeded by a subsequent article that discussed the superintegrability and exact solvability of the Dunkl-Coulomb problem in two-dimensional space \cite{Vincent3}. Following that work, Salazar-Ram\'irez et al. constructed their coherent states in \cite{Salazar2017, Salazar2018}. Then, Ghazouani et al. addressed the Dunkl-Coulomb problem in three- and arbitrary dimensional cases in \cite{Ghazouani2020, Ghazouani2021}. In the relativistic regime, the Dunkl operator was integrated into the Dirac equation to form the Dunkl-Dirac oscillator in one and two dimensions by Sargolzaeipor et al. and  Mota et al., respectively in \cite{Sargol2018, Mota20181}. In subsequent studies, Gullien et al. showed that algebraic methods could also be used to determine the energy spectrum of the one-dimensional Dunkl-Dirac oscillator \cite{Ojeda2020}. Later, two authors of this manuscript discussed the relation of the Dunk-Dirac oscillator to the Anti-Jaynes-Cummings Model \cite{Bilel20222}. Like the Dirac equation, the Klein-Gordon equation also coupled to an oscillator or Coulomb potential was solved using the Dunkl operator in place of partial derivatives in two and three dimensions, respectively \cite{Mota20212, Bilel20221}. Besides the Landau levels were obtained by solving the Dunkl-Klein-Gordon oscillator coupled to an external magnetic field in two dimensions, either with the $su(1,1)$ Lie algebra or analytically, as detailed in reference \cite{Mota20211}. Recently, Schulze-Halberg obtained the closed-form solutions of the Dunkl–Klein–Gordon equation for two inverse power-law interactions \cite{Schulze20243}.  An additional intriguing relativistic oscillator, the Duffin-Kemmer-Petiau (DKP) oscillator, which addresses the dynamics of both spin-0 and spin-1 particles simultaneously, was first studied within the Dunkl formalism by Merad et al \cite{Merad2021}. Subsequently, the solutions of the  DKP equation with the Coulomb \cite{Merad2022} and step potentials \cite{Askari2023} were examined. More recently, the Dunkl-Pauli equation in the presence of an external magnetic field has been studied by the two authors of this manuscript \cite{Bouguerne2024}. Several authors have put forth a number of generalizations of the Dunkl derivative, which would be employed to achieve a superior alignment between the theoretical modeling and the experimental results \cite{Dong2021, Halberg2022, Samira2022, Mota20221, Dong2023, Rouabhia2023, Mota20241, Mota20242}.

A review of the literature reveals that the majority of research conducted in the field of physics utilizing the Dunkl formalism has been confined to low-dimensional systems. This paper aims to address this gap in the literature and to investigate the impact of higher dimensions on the Dunkl formalism by analyzing the Schrödinger equation in d dimensions. To achieve this, we examine two fundamental interactions in physics: the harmonic oscillator and the Coulomb potential. The manuscript is constructed in the following manner: In Section \ref{sec2}, we generalize the Dunkl derivative and its formalism to d-dimensions. Then, the Dunkl-Schr\"odinger equation is solved with an isotropic harmonic oscillator in Cartesian coordinates. Subsequently, the polar coordinates in d-dimensions are introduced and the Dunkl formalism and Schr\"odinger equation are adapted. Following this, the angular part solution is discussed and the radial equation for the harmonic and pseudoharmonic potentials is investigated in Section \ref{sec3}. The Coulomb potential solutions are then obtained in Section  \ref{sec4}. Finally, the manuscript is concluded with a brief conclusion section.


\section{Dunkl Oscillator Hamiltonian in d-dimensions} \label{sec2}

The d-dimensional isotropic Dunkl oscillator model may be regarded as an extension of the original model, in which the standard derivatives are substituted with Dunkl derivatives, thereby introducing supplementary singular terms into the potential. In this context, the Hamiltonian of the d-dimensional isotropic Dunkl oscillator model is defined by
\begin{equation}
H=-\frac{\hbar ^{2}}{2m}\left( D_{1}^{2}+D_{2}^{2}+\cdots +D_{d}^{2}\right) +%
\frac{m\omega ^{2}}{2}\left( x_{1}^{2}+x_{2}^{2}+\cdots +x_{d}^{2}\right) ,
\label{e1}
\end{equation}
where $D_{j}$ is the Dunkl derivative \cite{Dunkl1989, Genest20131}%
\begin{equation}
D_{j}=\frac{\partial }{\partial x_{j}}+\frac{\mu _{j}}{x_{j}}\left( \mathbf{1}-R_{j}\right), \qquad \text{for} \quad  j=1,2,3,\cdots ,d \,,
\end{equation}
with the Wigner deformation constant $\mu _{j}$ and the identity operator $\mathbf{1}$. Here, the reflection operator, $R_{j}$, acts on the eigenfunction as follows:
\begin{equation}
R_{j}f\left( x_{1},x_{2},\cdots ,x_{j},x_{j+1},\cdots ,x_{d}\right) =f\left(
x_{1},x_{2},\cdots ,-x_{j},x_{j+1},\cdots ,x_{d}\right),
\end{equation}%
thus,
\begin{equation}
    R_{j}^{2}=\mathbf{1}.
\end{equation}
In order to preserve the symmetry in the full Hamiltonian, we impose the condition that all deformation coefficients are equal
\begin{equation}
\mu _{1}=\mu _{2}=\cdots =\mu _{d}.
\end{equation}%

\subsection{Solution with the Cartesian coordinates}

In the d-dimensional Cartesian coordinates, the Schr\"{o}dinger equation associated to the Hamiltonian given in Eq. \eqref{e1} can be expressed as follows:  
\begin{eqnarray}
\left[ -\frac{\hbar ^{2}}{2m}\left( D_{1}^{2}+D_{2}^{2}+\cdots
+D_{d}^{2}\right) +\frac{m\omega ^{2}}{2}\left( x_{1}^{2}+x_{2}^{2}+\cdots
+x_{d}^{2}\right) \right] \psi \left( x_{1},x_{2},\cdots ,x_{d}\right) =%
\mathcal{E}\psi \left( x_{1},x_{2},\cdots ,x_{d}\right) .  \label{e2}
\end{eqnarray}
The most straightforward solution to this differential equation can be obtained through the separation of variables method. In this case, the total Hamiltonian can be expressed as the sum of $d$ sub-Hamiltonian terms
\begin{equation}
    H=H_{1}+H_{2}+\cdots +H_{d},
\end{equation}
where each sub-Hamiltonian term 
\begin{equation}
H_{j}=-\frac{\hbar ^{2}}{2m}D_{j}^{2}+\frac{m\omega ^{2}}{2}x_{j}^{2}, 
\end{equation}
corresponds to a simultaneous one-dimensional Dunkl oscillator  
\begin{eqnarray}
    H_{j} \psi \left( x_{j}\right)=\varepsilon _{j}\psi \left( x_{j}\right).
\end{eqnarray}
Here, the total eigenfunction is taken as the multiplication of each separated one-dimensional solutions 
\begin{equation}
\psi \left( x_{1},x_{2},\cdots ,x_{d}\right) =\psi \left( x_{1}\right) \psi \left( x_{2}\right) \cdots \psi \left( x_{d}\right),  \label{e12}
\end{equation}%
with the total eigenvalue 
\begin{eqnarray}
    \mathcal{E}=\varepsilon _{1}+\varepsilon _{2}+\cdots +\varepsilon _{d}.  \label{en}
\end{eqnarray}
In the literature the solution of the  non-relativistic one-dimensional Dunkl oscillator
\begin{equation}
\left[ -\frac{\hbar ^{2}}{2m}\left( \frac{\partial ^{2}}{\partial x_{j}^{2}}+%
\frac{2\mu _{j}}{x_{j}}\frac{\partial }{\partial x_{j}}-\frac{\mu _{j}}{%
x_{j}^{2}}\left( \mathbf{1}-R_{j}\right) \right) +\frac{m\omega ^{2}}{2}%
x_{j}^{2}\right] \psi \left( x_{j}\right) =\varepsilon _{j}\psi \left(
x_{j}\right) ,  \label{e4}
\end{equation}%
were studied extensively. In order to avoid unnecessary repetition, we present a brief overview of the solution here. Given that the $H_{j}$ and $R_{j}$ operators commute, we select eigenfunctions with a specific parity
\begin{eqnarray}
 R_{j}\psi \left(x_{j}\right) =s_{j}\psi \left( x_{j}\right), 
 \end{eqnarray}
where $s_{j}=\pm 1.$ Then, we express the parity-dependent solutions:
\begin{itemize}
\item For the case $s_{j}=+1$, we have 
\begin{equation}
 R_{j}\psi ^{+}\left(x_{j}\right) =\psi ^{+}\left( x_{j}\right),    
\end{equation}
thus, 
\begin{equation}
\left[ -\frac{\hbar ^{2}}{2m}\left( \frac{\partial ^{2}}{\partial x_{j}^{2}}+%
\frac{2\mu _{j}}{x_{j}}\frac{\partial }{\partial x_{j}}\right) +\frac{%
m\omega ^{2}}{2}x_{j}^{2}\right] \psi^{+} \left( x_{j}\right) =\varepsilon
_{j}^{+}\psi ^{+}\left( x_{j}\right) .  \label{18}
\end{equation}%
Then, the complete solution of Eq. \eqref{18} stands in terms of the
associated Laguerre polynomials \cite{Gradshteyn}
\begin{equation}
\psi ^{+}\left( x_{j}\right) =C_{j}^{+}e^{-\frac{m\omega }{2\hbar }x^{2}} \mathbf{L}_{n_{j}}^{\mu _{j}-1/2}\left( \frac{m\omega }{\hbar }x^{2}\right), 
\end{equation}%
with eigenvalues
\begin{equation}
\varepsilon _{j}^{+}=\hbar \omega \left( 2n_{j}+\mu _{j}+1/2\right), \qquad n_{j}\in \left\{ 1,2,3,\cdots \right\} .
\end{equation}

\item For the case $s_{j}=-1$, we have 
\begin{equation}
R_{j}\psi ^{-}\left(
x_{j}\right) =-\psi ^{-}\left( x_{j}\right),     
\end{equation}
so that
\begin{equation}
\left[ -\frac{\hbar ^{2}}{2m}\left( \frac{\partial ^{2}}{\partial x_{j}^{2}}+%
\frac{2\mu _{j}}{x_{j}}\frac{\partial }{\partial x_{j}}-\frac{2\mu _{j}}{%
x_{j}^{2}}\right) +\frac{m\omega ^{2}}{2}x_{j}^{2}\right] \psi ^{-}\left(
x_{j}\right) =\varepsilon _{j}^{-}\psi ^{-}\left( x_{j}\right) .  \label{12}
\end{equation}%
Then, the odd solution appears as
\begin{equation}
\psi ^{-}\left( x_{j}\right) =C_{j}^{-}e^{-\frac{m\omega }{2\hbar }x^{2}} x \text{ }\mathbf{L}_{n_{j}}^{\mu _{j}+1/2}\left( \frac{m\omega }{\hbar }x^{2}\right),
\end{equation}
with eigenvalues
\begin{equation}
\varepsilon _{j}^{-}=\hbar \omega \left( 2n_{j}+\mu _{j}+3/2\right), \qquad n_{j}\in \left\{ 1,2,3,\cdots \right\} .
\end{equation}
\end{itemize}
The conjunction of the outcomes of the two subcases enables the general solution of Eq. \eqref{e4} to be expressed as follows:
\begin{eqnarray}
\psi ^{s_{j}}\left( x_{j}\right) &=&C_{j}^{s_{j}}e^{-\frac{m\omega }{2\hbar }x^{2}}x^{\frac{1-s_{j}}{2}}\text{ }\mathbf{L}_{n_{j}}^{\mu_{j}-s_{j}/2}\left( \frac{m\omega }{\hbar }x^{2}\right) , \\ 
\varepsilon _{j}^{s_{j}}&=&\hbar \omega \left( 2n_{j}+\mu_{j}+1-s_{j}/2\right) .
\end{eqnarray}%
The total energy eigenvalue expression is determined by substituting the specific $\varepsilon _{j}^{s_{j}}$ expressions defined in Eq. \eqref{en}. This calculation yields our final result 
\begin{equation}
\mathcal{E}_{n}=\hbar \omega \left[ 2n+d+\left( \mu _{1}+\mu _{2}+\cdots
+\mu _{d}\right) -\left( s_{1}+s_{2}+\cdots +s_{d}\right) /2\right] ,
\end{equation}
where $n=n_{1}+n_{2}+\cdots +n_{d}.$ Our results indicate that the total energy eigenvalue function of the Dunkl-oscillator is affected by the quantum numbers "$n$" as well as the Wigner parameters and parities. The highest contribution from the additional terms occurs when $s_{1}=s_{2}=\cdots =s_{d}=-1$, whereas the lowest contribution is achieved when $s_{1}=s_{2}=\cdots =s_{d}=+1.$

\subsection{Solution with the polar coordinates}

We now examine the Dunkl-Schr\"{o}dinger equation using d-dimensional polar coordinates. To do so, we will employ a set of coordinates with one spatial, $r$, and $d-1$ angular, $\theta _{j}$, components. In this case, the reflection operators will act as follows: 
\begin{eqnarray}
R_{1}f\left( r,\theta _{1},\theta _{2},\cdots ,\theta _{j},\cdots ,\theta
_{d-1}\right) &=& f\left( r,\pi -\theta _{1},\theta _{2},\cdots ,\theta
_{j},\cdots ,\theta _{d-1}\right), \nonumber \\ 
R_{2}f\left( r,\theta _{1},\theta _{2},\cdots ,\theta _{j},\cdots ,\theta
_{d-1}\right) &=&f\left( r,-\theta _{1},\theta _{2},\cdots ,\theta _{j},\cdots
,\theta _{d-1}\right), \nonumber \\ 
&\vdots& \nonumber  \\ 
R_{j}f\left( r,\theta _{1},\theta _{2},\cdots ,\theta _{j},\cdots ,\theta
_{d-1}\right) &=& f\left( r,\theta _{1},\theta _{2},\cdots ,\pi -\theta
_{j},\cdots ,\theta _{d-1}\right), \\ 
&\vdots& \nonumber \\ 
R_{d}f\left( r,\theta _{1},\theta _{2},\cdots ,\theta _{j},\cdots ,\theta
_{d-1}\right) &=&f\left( r,\theta _{1},\theta _{2},\cdots ,\theta _{j},\cdots
,\pi -\theta _{d-1}\right). \nonumber
\end{eqnarray}
Let us start by defining the relationships between the Cartesian coordinates $x_{j}$ and the d-dimensional polar coordinates \cite{Louck1960, Dong}:
\begin{eqnarray} \label{e5}
x_{1}&=& r \cos \theta _{1}\sin \theta _{2}\sin \theta _{3}\cdots \sin \theta _{d-1}, \nonumber\\ 
x_{2}&=& r \sin \theta _{1}\sin \theta _{2}\sin \theta _{3}\cdots \sin \theta _{d-1}, \nonumber\\ 
x_{3}&=& r \cos \theta _{2}\sin \theta _{3}\sin \theta _{4}\cdots \sin \theta _{d-1}, \nonumber\\ 
&\vdots& \nonumber \\ 
x_{j}&=&r\cos \theta _{j-1}\sin \theta _{j}\sin \theta _{j+1}\cdots \sin \theta _{d-1},  \\ 
&\vdots&  \nonumber\\ 
x_{d-1}&=& r\cos \theta _{d-2}\sin \theta _{d-1}, \nonumber \\ 
x_{d} &=& r\cos \theta _{d-1},  \qquad \qquad  \qquad \text{for} \,\, d=3,4,5,\cdots,d-1. \nonumber  
\end{eqnarray}
Here, the range of the spatial component is $r\in \left]0,+\infty \right[ ,$ while the angular components ranges are $0\leq \theta _{1}\leq 2\pi $, and $0\leq \theta
_{j}\leq \pi $ for $j=2,3,\cdots, d-1$. In this notation the sum of the squares of Eq. \eqref{e5} gives
\begin{equation}
\sum_{j=1}^{d}x_{j}^{2}=r^{2}.
\end{equation}
We now consider the Dunkl derivative. Taking the square of the Dunkl derivative in d-dimensions, we get
\begin{equation}
\sum_{j=1}^{d}D_{j}^{2}=\Delta +\sum_{j=1}^{d}\left( \frac{2\mu _{j}}{x_{j}}%
\frac{\partial }{\partial x_{j}}-\frac{\mu _{j}}{x_{j}^{2}}\left( \mathbf{1}%
-R_{j}\right) \right).
\end{equation}
Here, the first term is the conventional Laplacian operator and it can be expressed in d-dimensional polar coordinates with
\begin{eqnarray}
\Delta &=&\frac{\partial ^{2}}{\partial r^{2}}+\frac{d-1}{r}\frac{\partial }{%
\partial r}+\frac{1}{r^{2}}\sum_{j=1}^{d-2}\frac{1}{\sin ^{2}\theta
_{j+1}\sin ^{2}\theta _{j+2}\cdots \sin ^{2}\theta _{d-1}}\bigg[\frac{%
\partial ^{2}}{\partial \theta _{j}^{2}}+\left( j-1\right) \tan \theta _{j}\frac{\partial }{\partial \theta _{j}}\bigg]  \notag \\
&&+\frac{1}{r^{2}}\bigg[ \frac{1}{\sin ^{d-2}\theta _{d-1}}\frac{\partial }{\partial \theta _{d-1}}\sin ^{d-2}\theta _{d-1}\frac{\partial }{\partial \theta _{d-1}}\bigg] .
\end{eqnarray}%
In polar coordinates, for $d\geq 3$ we have
\begin{equation}
\frac{\partial }{\partial x_{j}}=\sum_{k=1}^{d}\frac{\partial \theta _{k}}{%
\partial x_{j}}\frac{\partial }{\partial \theta _{k}}=\sum_{k=1}^{d}\left( 
\frac{1}{h_{j}^{2}}\frac{\partial x_{k}}{\partial \theta _{j}}\right) \frac{%
\partial }{\partial \theta _{j}},
\end{equation}%
where%
\begin{eqnarray}
h_{0}&=& 1, \nonumber\\ 
h_{1}&=& r \sin \theta _{2}\sin \theta _{3}\cdots \sin \theta _{d-1}, \nonumber\\ 
h_{2}&=& r\sin \theta _{3}\sin \theta _{4}\cdots \sin \theta _{d-1}, \nonumber\\ 
h_{3}&=& r\sin \theta _{4}\sin \theta _{5}\cdots \sin \theta _{d-1}, \nonumber\\ 
&\vdots& \nonumber \\ 
h_{j}&=& r \sin \theta _{j+1}\sin \theta _{j+2}\cdots \sin \theta _{d-1}, \\ 
&\vdots& \nonumber \\ 
h_{d-2}&=& r\sin \theta _{d-1},  \nonumber \\
h_{d-1}&=& r, \nonumber
\end{eqnarray}
and
\begin{eqnarray}
    h=r^{d-1}\sin \theta _{2} \sin^2  \theta _{3} \sin^3  \theta _{4}\cdots \sin^{d-2}  \theta _{d-1}.
\end{eqnarray}
Then, for a particle that moves under a spherically symmetric potential $V\left( r\right),$ the Dunkl-Schr\"{o}dinger equation takes the form%
\begin{eqnarray}
&&\left[ \mathcal{A}_{r}+\frac{\hbar ^{2}}{2mr^{2}} \bigg( \frac{\mathcal{J}_{\theta _{1}}}{\sin ^{2}\theta _{2}\sin ^{2}\theta _{3}...\sin ^{2}\theta _{d-1}}+\frac{\mathcal{J}_{\theta _{2}}}{\sin ^{2}\theta _{3}...\sin ^{2}\theta _{d-1}}+\cdots +\mathcal{J} _{\theta _{d-1}}\bigg)\right] \psi \left( r,\theta _{1},\cdots ,\theta_{d-1}\right) \nonumber\\ 
&=& E\psi \left( r,\theta _{1},\cdots ,\theta _{d-1}\right) ,
\label{schro}
\end{eqnarray}
where%
\begin{equation}
\mathcal{A}_{r}=-\frac{\hbar ^{2}}{2m}\bigg[ \frac{\partial ^{2}}{\partial r^{2}}+\frac{d-1+2\left( \mu _{1}+\mu _{2}+\mu _{3}+...+\mu _{d}\right) }{r} \frac{\partial }{\partial r}\bigg] +V\left( r\right). \label{zam1}
\end{equation}
Here, the angular operators $\mathcal{J}_{\theta _{1}},\mathcal{J}_{\theta _{2}},\cdots ,\mathcal{J}_{\theta _{d-1}}$, can be expressed in terms of polar coordinates as follows:
\small
\begin{eqnarray}
\mathcal{J}_{\theta _{1}}&=&-\frac{\partial ^{2}}{\partial \theta _{1}^{2}}+\frac{2\left( \mu _{1}\tan \theta _{1}-\mu _{2}\cot \theta _{1}\right) }{r^{2}}\frac{\partial }{\partial \theta _{1}}+\frac{\mu _{1}}{\cos ^{2}\theta_{1}}\left( 1-R_{1}\right) +\frac{\mu _{2}\left( 1-R_{2}\right) }{\sin^{2}\theta _{1}}, \nonumber \\ 
\mathcal{J}_{\theta _{2}}&=&-\frac{\partial ^{2}}{\partial \theta _{2}^{2}}-\bigg[ \left( 1+2\left( \mu _{1}+\mu _{2}\right) \right) \cot \theta_{2}-2\mu _{3}\tan \theta _{2}\bigg] \frac{\partial }{\partial \theta _{2}}+\frac{\mu _{3}\left( 1-R_{3}\right) }{\cos ^{2}\theta _{2}}, \nonumber\\ 
\mathcal{J}_{\theta _{3}}&=&-\frac{\partial ^{2}}{\partial \theta _{3}^{2}}-
\bigg[ \left( 2+2\left( \mu _{1}+\mu _{2}+\mu _{3}\right) \right) \cot
\theta _{3}-2\mu _{4}\tan \theta _{3}\bigg] \frac{\partial }{\partial
\theta _{3}}+\frac{\mu _{4}\left( 1-R_{4}\right) }{\cos ^{2}\theta _{3}},  \nonumber \\ 
&\vdots&  \nonumber  \\ 
\mathcal{J}_{\theta _{d-2}}&=&-\frac{\partial ^{2}}{\partial \theta _{d-2}^{2}}%
-\bigg[ \left( \left( d-3\right) +2\left( \mu _{1}+\mu _{2}+...+\mu
_{d-2}\right) \right) \cot \theta _{d-2}-2\mu _{d-1}\tan \theta _{d-2}\bigg]\frac{\partial }{\partial \theta _{d-2}}+\frac{\mu _{d-1}\left(
1-R_{d-1}\right) }{\cos ^{2}\theta _{d-2}}, \nonumber \\ 
\mathcal{J}_{\theta _{d-1}}&=&\frac{\partial ^{2}}{\partial \theta _{d-1}^{2}}+\bigg[ \left( \left( d-2\right) +2\left( \mu _{1}+\mu _{2}+...+\mu_{d-1}\right) \right) \cot \theta _{d-1}-2\mu _{d}\tan \theta _{d-1}\bigg]\frac{\partial }{\partial \theta _{d-1}}+\frac{\mu _{d}\left( 1-R_{d}\right)}{\cos ^{2}\theta _{d-1}}. 
\end{eqnarray}
\normalsize
We observe that each of these operators depends only on one of the angles of $\theta _{1},\cdots ,\theta _{d-1}$, so their eigenstates can be denoted by $\Theta _{1}\left( \theta _{1}\right) ,\Theta _{2}\left( \theta _{2}\right) ,\cdots ,\Theta _{d-1}\left( \theta
_{d-1}\right) $. 
Now, we assume the total wave function in the following form
\begin{equation}
 \psi \left( r,\theta _{1},\cdots ,\theta _{d-1}\right) =U\left(
r\right) \Theta _{1}\left( \theta _{1}\right) \Theta _{2}\left( \theta
_{2}\right) \cdots \Theta _{d-1}\left( \theta _{d-1}\right).    
\end{equation}
This allows us to transform the Schr\"{o}dinger equation, given in Eq. \eqref{schro}, into a set of ordinary differential equations with one radial equation 
\begin{equation}
\left[ -\frac{\hbar ^{2}}{2m}\left( \frac{\partial ^{2}}{\partial r^{2}}+\frac{d-1+2\left( \mu _{1}+\mu _{2}+\mu _{3}+\cdots +\mu _{d}\right) }{r}\frac{\partial }{\partial r}\right) +V\left( r\right) +\frac{\hbar^{2}\varpi ^{2}}{2mr^{2}}\right] U\left( r\right) =EU\left( r\right) ,
\end{equation}
and $d-1$ angular equations
\begin{eqnarray}
\mathcal{J}_{\theta _{1}}\Theta _{1}\left( \theta _{1}\right) &=&\lambda_{1}^{2}\Theta _{1}\left( \theta _{1}\right) , \\
\left( \mathcal{J}_{\theta _{2}}+\frac{\lambda _{1}^{2}}{\sin ^{2}\theta _{2}}\right) \Theta _{2}\left( \theta _{2}\right) &=&\lambda _{2}^{2}\Theta_{2}\left( \theta _{2}\right) , \\
\left( \mathcal{J}_{\theta _{3}}+\frac{\lambda _{2}^{2}}{\sin ^{2}\theta _{3}}\right) \Theta _{3}\left( \theta _{3}\right) &=&\lambda _{3}^{2}\Theta_{d-1}\left( \theta _{d-1}\right) , \\
&\vdots& \nonumber \\
\left( \mathcal{J}_{\theta _{d-1}}+\frac{\lambda _{d-2}^{2}}{\sin ^{2}\theta_{d-1}}\right) \Theta _{d-1}\left( \theta _{d-1}\right) &=&\varpi ^{2}\Theta_{d-1}\left( \theta _{d-1}\right) .
\end{eqnarray}
where $\lambda _{1},\lambda _{2},...,\lambda _{d-2}$ and $\varpi $ are the separation constants.

The solution of the first angular component,  $\Theta _{1}\left( \theta_{1}\right) $, is characterized by the eigenvalues $s_{1}=\pm $, $s_{2}=\pm$ of the reflection operators $R_{1}$ and $R_{2}$,  respectively.
\begin{equation}
\Theta _{1}^{s_{1},s_{2}}\left( \theta _{1}\right) =i_{\ell _{1}}\cos
^{e_{1}}\theta _{1}\sin ^{e_{2}}\theta _{1}\mathbf{P}_{\ell _{1}-\left(
e_{1}+e_{2}\right) /2}^{\left( \mu _{2}+e_{2}-1/2;\mu _{1}+e_{1}-1/2\right)
}\left( \cos 2\theta _{1}\right) .  \label{sol1}
\end{equation}
Here, $i_{\ell _{1}}$is normalization constant, $\mathbf{P}_{n}^{\left(\alpha ;\beta \right) }\left( x\right) $ is Jacobi polynomials and 
\begin{equation}
e_{j}=\left\{ 
\begin{array}{ll}
1, &\text{ \ \ \ if } s_{j}=-1, \\ 
0, &\text{ \ \ \  if }s_{j}=1.
\end{array}%
\right. 
\end{equation}
The solution of Eq. \eqref{sol1} corresponds to the eigenvalue 
\begin{eqnarray}
 \lambda_{1}^{2}=4\ell _{1}\left( \ell _{1}+\mu _{1}+\mu _{2}\right).   
\end{eqnarray}
It is important to mention that when $s_{1}\cdot s_{2}=-1$, $\ell _{1}$ takes on values that are positive half-integers (e.g.,1/2,3/2,...).  Conversely, when $s_{1}=s_{2}=+1$, $\ell _{1}$ is a non-negative integer. Note also that in
the special case where $\ell _{1}=0$, only the $s_{1}=s_{2}=+1$ state exists. 

The angular solution of the second angular component,  $\Theta _{2}\left( \theta _{2}\right)$, are labeled by the eigenvalue $s_{3}=\pm $ of the reflection operator $R_{3}$. One has
\begin{equation}
\Theta _{2}^{s_{3}}\left( \theta _{2}\right) =i_{\ell _{2}}\cos^{e_{3}}\theta _{1}\sin ^{2\ell _{1}}\theta _{2}\mathbf{P}_{\ell _{2}-\frac{e_{3}}{2}}^{\left( 2\ell _{1}+\mu _{1}+\mu _{2};\mu _{3}+e-1/2\right)}\left( \cos 2\theta _{2}\right) .
\end{equation}
For the case where $s_{3}=1$, $\ell _{2}$ takes on values that are non-negative integers. In contrast, when $s_{3}=-1$, $\ell _{2}$ receives values that are positive half-integers. The separation constant is assigned the value $\lambda _{2}$, which is equal to%
\begin{equation}
\lambda _{2}^{2}=4\left( \ell _{2}+\ell _{1}\right) \left( \ell _{2}+\ell
_{1}+\mu _{1}+\mu _{2}+\mu _{3}+1/2\right) .
\end{equation}%
Following the same procedure, one can derive the whole set of polar coordinate solutions of the
remaining angular parts, $\Theta _{j}^{s_{j+1}}\left( \theta _{j}\right) $.  Here, we tabulate them in Table \ref{table:1} and their eigenvalues in Table \ref{table:2}.
\begin{table}[H]
\centering
\begin{tabular}{||l|l||}
\hline
$j$ & $\Theta _{j}^{s_{j+1}}\left( \theta _{j}\right) $ \\ \hline
$3$ & $\cos ^{e_{4}}\theta _{3}\sin ^{2\left( \ell _{2}+\ell _{1}\right)
}\theta _{3}\mathbf{P}_{\ell _{3}-\frac{e_{4}}{2}}^{\left( 1/2+2\left( \ell
_{2}+\ell _{1}\right) +\mu _{1}+\mu _{2}+\mu _{3},\mu _{4}+e_{4}-1/2\right)
}\left( \cos 2\theta _{3}\right) $ \\ \hline
$4$ & $\cos ^{e_{5}}\theta _{2}\sin ^{2\left( \ell _{2}+\ell _{1}+\ell
_{3}\right) }\theta _{2}\mathbf{P}_{\ell _{4}-\frac{e_{5}}{2}}^{\left(
1+2\left( \ell _{1}+\ell _{2}+\ell _{3}\right) +\mu _{1}+\cdots +\mu
_{4},\mu _{5}+e_{5}-1/2\right) }\left( \cos 2\theta _{4}\right) $ \\ \hline
$5$ & $\cos ^{e_{6}}\theta _{5}\sin ^{2\left( \ell _{1}+\cdots +\ell
_{4}\right) }\theta _{5}\mathbf{P}_{\ell _{5}-\frac{e_{6}}{2}}^{\left(
3/2+2\left( \ell _{1}+\cdots +\ell _{4}\right) +\mu _{1}+\cdots +\mu
_{5},\mu _{6}+e_{6}-1/2\right) }\left( \cos 2\theta _{5}\right) $ \\ \hline
$6$ & $\cos ^{e_{7}}\theta _{6}\sin ^{2\left( \ell _{1}+\cdots +\ell
_{5}\right) }\theta _{6}\mathbf{P}_{\ell _{6}-\frac{e_{7}}{2}}^{\left(
2+2\left( \ell _{1}+..+\ell _{5}\right) +\mu _{1}+\cdots +\mu _{5},\mu
_{7}+e_{7}-1/2\right) }\left( \cos 2\theta _{6}\right) $ \\ \hline
$\vdots $ & $\vdots $ \\ \hline
$k$ & $\cos ^{e_{k+1}}\theta _{k}\sin ^{2\left( \ell _{1}+\cdots +\ell
_{k-1}\right) }\theta _{k}\mathbf{P}_{\ell _{k}-\frac{e_{k+1}}{2}}^{\left( 
\frac{k-2}{2}+2\left( \ell _{1}+..+\ell _{k-1}\right) +\mu _{1}+\cdots +\mu
_{k},\mu _{k+1}+e_{k+1}-1/2\right) }\left( \cos 2\theta _{k}\right) $ \\ 
\hline
\end{tabular}%
\caption{First few angular solutions $\Theta _{j}^{s_{j+1}}\left( \protect%
\theta _{j}\right) $.}
\label{table:1}
\end{table}

\begin{table}[H]
\centering
\begin{tabular}{||l|l||}
\hline
$j$ & $\lambda _{j}^{2}$ \\ \hline
$3$ & $4\left( \ell _{1}+\ell _{2}+\ell _{3}\right) \left( \ell _{1}+\ell
_{2}+\ell _{3}+\mu _{1}+\cdots +\mu _{4}+1\right) $ \\ \hline
$4$ & $4\left( \ell _{1}+\ell _{2}+\ell _{3}+\ell _{4}\right) \left( \ell
_{1}+\cdots +\ell _{4}+\mu _{1}+\cdots +\mu _{5}+3/2\right) $ \\ \hline
$5$ & $4\left( \ell _{1}+\ell _{2}+\cdots +\ell _{5}\right) \left( \ell _{1}+\cdots
+\ell _{5}+\mu _{1}+\cdots +\mu _{6}+2\right) $ \\ \hline
$6$ & $4\left( \ell _{1}+\ell _{2}+\cdots +\ell _{6}\right) \left( \ell
_{1}+\ell _{2}+\cdots +\ell _{6}+\mu _{1}+\cdots +\mu _{7}+5/2\right) $ \\ 
\hline
$\vdots $ & $\vdots $ \\ \hline
$k$ & $4\left( \ell _{1}+\ell _{2}+\cdots +\ell _{k}\right) \left( \ell
_{1}+\ell _{2}+\cdots +\ell _{k}+\mu _{1}+\cdots +\mu _{k+1}+\frac{k-1}{2}%
\right) $ \\ \hline
\end{tabular}%
\caption{First few separation constant $\protect\lambda _{j}^{2}$.}
\label{table:2}
\end{table}
Next, we look for an exact solution to the radial equation. 
\begin{equation}
\left[ \frac{d^{2}}{dr^{2}}+\frac{d-1+2\left( \mu _{1}+\mu _{2}+\mu
_{3}+\cdots +\mu _{d}\right) }{r}\frac{d}{dr}-\frac{2m}{\hbar ^{2}}V\left(
r\right) -\frac{\varpi ^{2}}{r^{2}}+\frac{2mE}{\hbar ^{2}}\right] U\left(
r\right) =0.  \label{rad}
\end{equation}
Here, the separation constant $\varpi ^{2}$ is given by 
\begin{equation}
\varpi ^{2}=4\left( \ell _{1}+\ell _{2}+\cdots +\ell _{d-1}\right) \left(
\ell _{1}+\ell _{2}+\cdots +\ell _{d-1}+\mu _{1}+\cdots +\mu _{d}+\frac{d-2}{%
2}\right) .
\end{equation}%
In the following sections, we will utilize Eq. \eqref{rad} to examine the Dunkl-Schr\"{o}dinger equation in two specific problems in physics: a harmonic oscillator and a Coulomb-like potential interactions.  The primary objective in examining these problems will be the determination of the energy eigenvalues and their associated eigenfunctions.

\section{Harmonic oscilator} \label{sec3}

\begin{figure}[htb!]
\begin{minipage}[t]{0.5\textwidth}
        \centering
        \includegraphics[width=\textwidth]{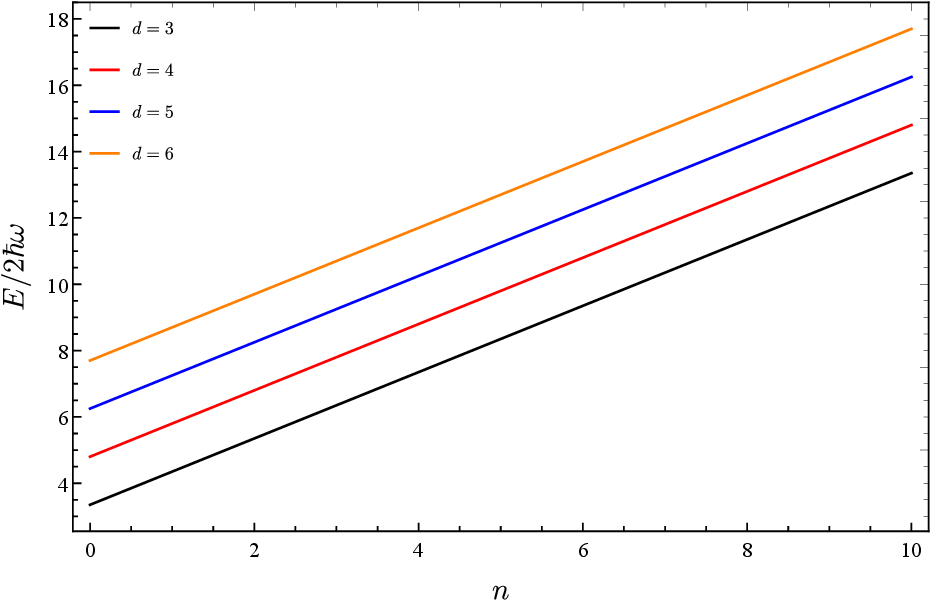}
        \label{fig:ta}
        \subcaption{$\mu_{i}=0.4$}
\end{minipage}
\begin{minipage}[t]{0.5\textwidth}
        \centering
        \includegraphics[width=\textwidth]{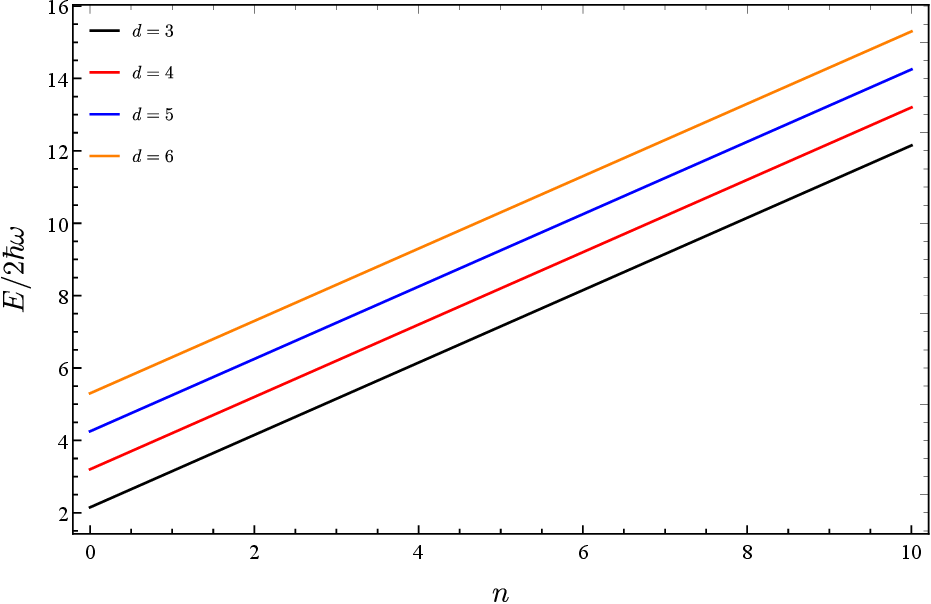}
       \label{fig:tb}
       \subcaption{$\mu_{i}=-0.4$}
   \end{minipage}
\caption{The energy eigenvalues vs. $n$ for various values of the spatial dimensions $d$.}
\label{fig:n}
\end{figure}

Within the Dunkl formalism, the energy spectrum and the eigenfunctions of the harmonic oscillator in three dimensions were determined in \cite{Bilel20221}. Here, we extend those solutions to d-dimensions. Using Eqs. \eqref{schro} and \eqref{zam1} we express the radial component of the Dunkl-Schr\"{o}dinger equation with the isotropic harmonic oscillator potential 
\begin{equation}
\left[ \frac{d^{2}}{dr^{2}}+\frac{d-1+2\left( \mu _{1}+\mu _{2}+\mu
_{3}+\cdots +\mu _{d}\right) }{r}\frac{d}{dr}-\frac{m^{2}\omega ^{2}}{\hbar
^{2}}r^{2}-\frac{\varpi ^{2}}{r^{2}}+\frac{2mE}{\hbar ^{2}}\right] U\left(
r\right) =0.  \label{pot1}
\end{equation}%
Here, $\omega $ denotes the vibration frequency of the particle. We then introduce a new variable 
\begin{equation}
\rho =\frac{m\omega }{\hbar }r^{2}, 
\end{equation}%
and transform Eq. \eqref{pot1} to
\begin{equation}
\left[ \rho \frac{d^{2}}{d\rho ^{2}}+\frac{1}{2}\left( d+2\left( \mu
_{1}+\mu _{2}+\mu _{3}+\cdots +\mu _{d}\right) \right) \frac{d}{d\rho }-%
\frac{\rho }{4}-\frac{\varpi ^{2}}{4\rho }+\frac{E}{2\hbar \omega }\right]
U\left( \rho \right) =0.
\end{equation}%
Following an examination of the radial function's behavior at the origin and infinity, we put forward the following solution to the wave function.
\begin{equation}
U\left( \rho \right) =e^{-\frac{\rho }{2}}\rho ^{\ell _{1}+\ell _{2}+\cdots
+\ell _{d-1}}\digamma \left( \rho \right) .
\end{equation}%
Subsequently, we obtain the following equation for $\digamma \left( \rho \right)$,  which mimics the confluent hypergeometric differential equation
\begin{equation}
\left\{ 
\begin{array}{c}
\rho \frac{d^{2}}{d\rho ^{2}}+\left[ \frac{d+2\left( \mu _{1}+\mu _{2}+\mu
_{3}+\cdots +\mu _{d}\right) }{2}+2\left( \ell _{1}+\ell _{2}+\cdots +\ell
_{d-1}\right) -\rho \right] \frac{d}{d\rho } \\ 
+\frac{E}{2\hbar \omega }-\left( \ell _{1}+\ell _{2}+\cdots +\ell
_{d-1}\right) -\frac{d+2\left( \mu _{1}+\mu _{2}+\mu _{3}+\cdots +\mu
_{d}\right) }{4}%
\end{array}%
\right\} \digamma \left( \rho \right) =0.
\end{equation}
So that we express the radial solution in the form of 
\begin{equation}
U\left( \rho \right) =\mathcal{C}\text{ \ }\rho ^{\ell _{1}+\ell _{2}+\cdots
+\ell _{d-1}}e^{-\frac{\rho }{2}}\mathbf{F}\left( a,b;\rho \right) ,
\label{pot3}
\end{equation}%
where $\mathcal{C}$ is the normalization factor to be determined and 
\begin{equation}
\left\{ 
\begin{array}{l}
a=\left( \ell _{1}+\ell _{2}+\cdots +\ell _{d-1}\right) +\frac{d+2\left( \mu
_{1}+\mu _{2}+\mu _{3}+\cdots +\mu _{d}\right) }{4}-\frac{E}{2\hbar \omega }
\\ 
b=\frac{d+2\left( \mu _{1}+\mu _{2}+\mu _{3}+\cdots +\mu _{d}\right) }{2}%
+2\left( \ell _{1}+\ell _{2}+\cdots +\ell _{d-1}\right)%
\end{array}%
\right. .
\end{equation}%
We now proceed to derive the energy spectrum. Given that the solutions are finite at infinity, we derive the general quantum condition from Eq. \eqref{pot3}
\begin{equation}
\left( \ell _{1}+\ell _{2}+\cdots +\ell _{d-1}\right) +\frac{d+2\left( \mu
_{1}+\mu _{2}+\mu _{3}+\cdots +\mu _{d}\right) }{4}-\frac{E}{2\hbar \omega }%
=-n,\text{ \ with\ \ }n\in \mathbb{N}.
\end{equation}%
Therefore, the energy spectrum reads:
\begin{equation}
E_{n,\ell _{1},...,\ell _{d-1},d}=2\hbar \omega \left( n+\ell _{1}+\ell
_{2}+\cdots +\ell _{d-1}+\frac{d+2\left( \mu _{1}+\mu _{2}+\mu _{3}+\cdots
+\mu _{d}\right) }{4}\right) .
\end{equation}
It is noteworthy that the expression of the energy spectrum can be compared to the special case when $d=3$, which yields the exact result for the three-dimensional Dunkl oscillator. Furthermore, it can be observed that the energy spectrum is influenced by a number of factors, including the Wigner deformation parameters $\mu _{i}$, the quantum numbers $n$ and $\ell _{i}$ as well as the spatial dimensions. In Fig. \ref{fig:n}, we plot the energy level as a function of quantum number $n$ for varying spatial dimensions $d$.

\begin{figure}[htb!]
\begin{minipage}[t]{0.33\textwidth}
        \centering
        \includegraphics[width=\textwidth]{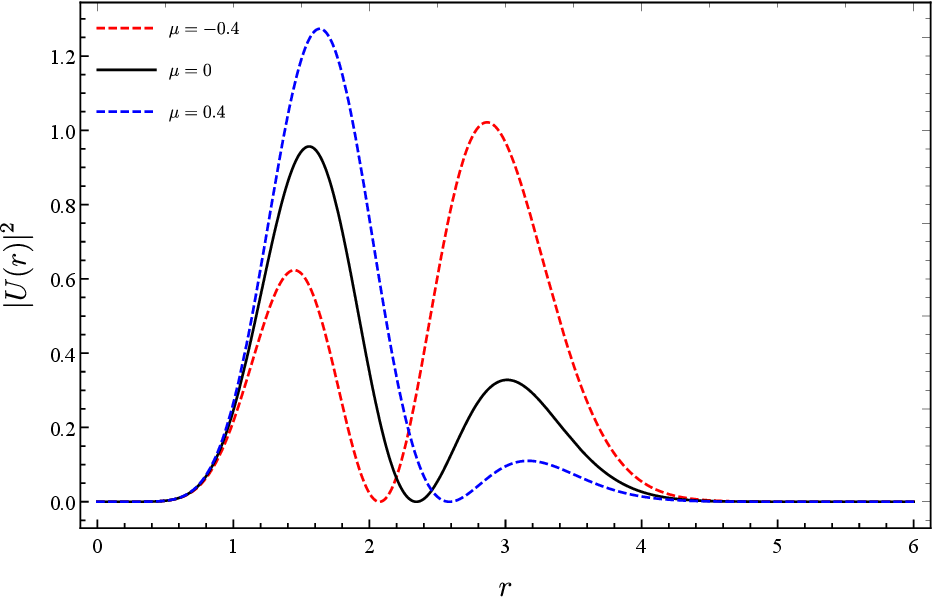}
         \subcaption{$d=3$ }\label{fig:phb}
\end{minipage}%
\begin{minipage}[t]{0.33\textwidth}
        \centering
        \includegraphics[width=\textwidth]{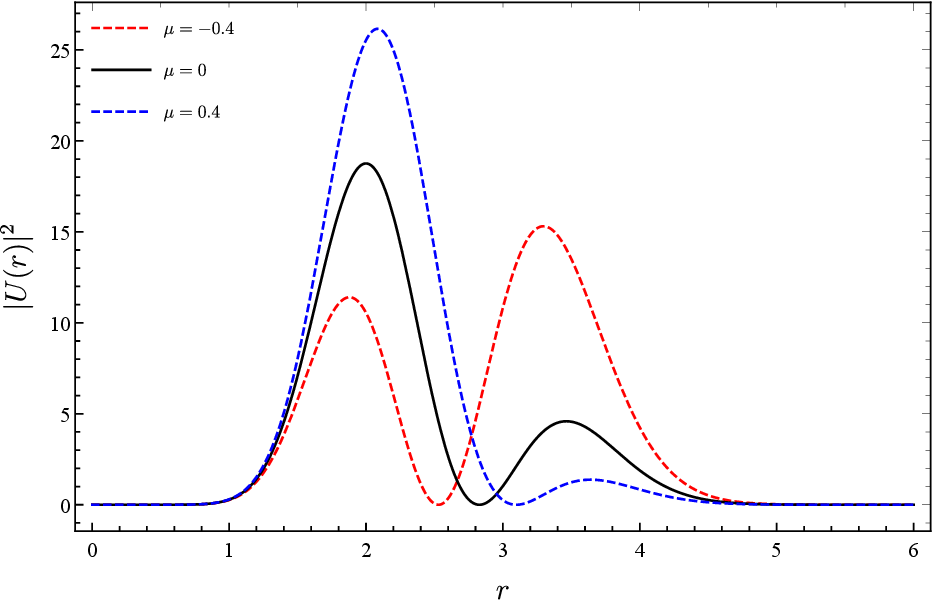}
       \subcaption{ $d=4$}\label{fig:phc}
   \end{minipage}%
\begin{minipage}[t]{0.33\textwidth}
        \centering
        \includegraphics[width=\textwidth]{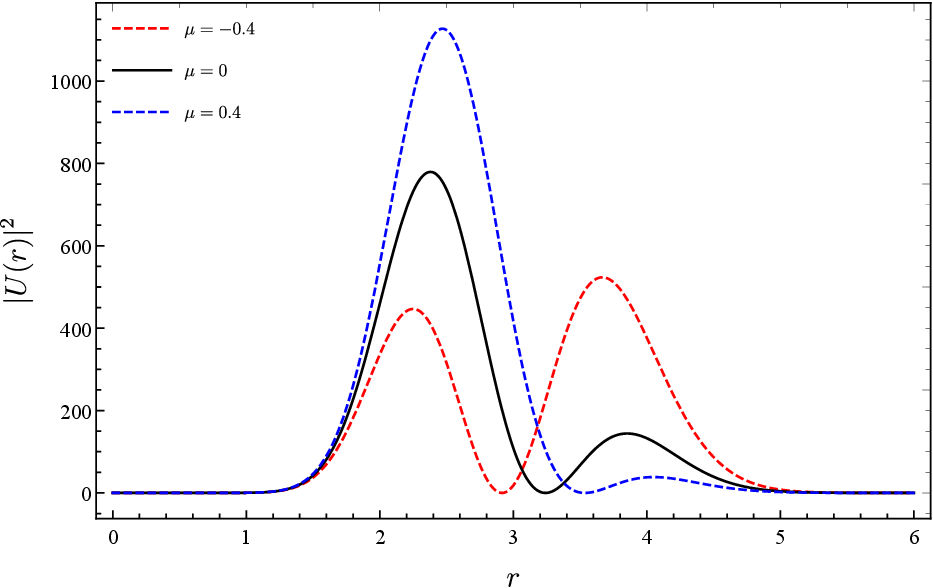}
         \subcaption{ $d=5$}\label{fig:phd}
   \end{minipage}
\caption{Reduced probability densities for various values of the Wigner parameter $n=1$ and $\ell_{i}=1$.}
\label{lafun}
\end{figure}

We observe that the energy level increases monotonically, and for a fixed value of $n$, the energy spectrum increases with the increase of the spatial dimensions parameter. It is worth to emphasize that in very high dimensions, the energy spectrum becomes 
\begin{equation}
E_{d}=\frac{d\hbar \omega }{2},
\end{equation}
and this result implies that the energy is almost independent of the quantum numbers. In Fig. \ref{lafun}, we present the reduced probability densities for three, four, and five dimensions.

It is evident that the Dunkl formalism modifies the probability densities. 

\subsection{Pseudoharmonic oscillator}
We now use our findings to examine the molecular pseudoharmonic oscillator (PHO) potential 
\begin{eqnarray}
 V\left( r\right) =D_{e}\left( \frac{r}{r_{e}}-\frac{r_{e}}{r}\right) ^{2},    
\end{eqnarray} 
where $D_{e}$  represents the dissociation energy and $r_{e}$ is the equilibrium internuclear distance \cite{Ikhdair2007}. The PHO is considered a crucial molecular potential, particularly significant in the study of diatomic molecules, and holds great importance in chemical physics, molecular physics, and other areas of physics research \cite{Oyewumi2012, Erkoc1988}. The PHO potential incorporates two key components: harmonic interactions and inverse square interactions. The Dunkl-Schr\"{o}dinger equation for PHO potential reads:
\begin{equation}
\left[ \frac{d^{2}}{dr^{2}}+\frac{d-1+2\left( \mu _{1}+\mu _{2}+\mu
_{3}+\cdots +\mu _{d}\right) }{r}\frac{d}{dr}-\frac{m^{2}\Omega ^{2}}{\hbar
^{2}}r^{2}-\frac{\delta ^{2}}{r^{2}}+\frac{2m}{\hbar ^{2}}\mathcal{E}\right]
U\left( r\right) =0,  \label{57}
\end{equation}%
where
\begin{eqnarray}
\delta ^{2}&=&\varpi ^{2}+\frac{2mD_{e}r_{e}^{2}}{\hbar ^{2}}, \\
\Omega ^{2}&=& \frac{4D_{e}}{mr_{e}^{2}}, \\
\mathcal{E}&=&E+2D_{e}.
\end{eqnarray}
It is evident that Eq. \eqref{57} exhibits a striking resemblance to Eq. \eqref{pot1}. Consequently, to avoid unnecessary repetition we omit the mathematical steps for the solutions and express the energy eigenvalue function as follows:
\begin{eqnarray}
E_{PHO} &=&-2D_{e}+4\hbar \sqrt{\frac{D_{e}}{mr_{e}^{2}}}\nonumber \\
&\times&\Big[n+\frac{1}{2} +\frac{1}{2}\sqrt{1+\left( \mu _{1}+\cdots +\mu
_{d}+d/2\right) \left( \mu _{1}+\cdots +\mu
_{d}+d/2-2\right) +\varpi ^{2}+\frac{2D_{e}mr_{e}^{2}}{\hbar ^{2}}}\Big].\label{102}
\end{eqnarray}
Eq. (\ref{102}) reveals the impact of the Dunkl algebra on the energy levels of the PHO. As highlighted in Ref. \cite{Ikhdair2007}, various characteristics of diatomic molecules can be analyzed using this equation. Furthermore, the free deformation parameter can be utilized to achieve precise alignment with experimental outcomes.

\section{Coulomb potential} \label{sec4}
In this section, we address another significant potential energy in physics. In one of our earlier papers \cite{Bilel20221}, we obtained an exact solution of the Dunkl-Schr\"{o}dinger equation with the Coulomb potential in three dimensions. In the present section, we revisit the same problem in d-dimensions with
\begin{equation}
\left[ \frac{d^{2}}{dr^{2}}+\frac{d-1+2\left( \mu _{1}+\mu _{2}+\mu
_{3}+\cdots +\mu _{d}\right) }{r}\frac{d}{dr}+\frac{2me^{2}}{\hbar ^{2}r}-%
\frac{\varpi ^{2}}{r^{2}}+\frac{2mE}{\hbar ^{2}}\right] U\left( r\right) =0.
\label{col1}
\end{equation}
Here, let us assume the solution to be in the following form
\begin{equation}
U\left( r\right) =r^{2\left( \ell _{1}+\ell _{2}+\cdots +\ell _{d-1}\right)
}e^{-\eta r}\Xi \left( r\right) ,
\end{equation}%
where $\eta =\frac{\sqrt{-2mE}}{\hbar }.$ Then, Eq. \eqref{col1} gives
\begin{equation}
\left\{ 
\begin{array}{c}
\frac{d^{2}}{dr^{2}}+\left( \frac{4\left( \ell _{1}+\ell _{2}+\cdots +\ell_{d-1}\right) +d-1+2\left( \mu _{1}+\mu _{2}+\mu _{3}+\cdots +\mu_{d}\right) }{r}-2\eta \right) \frac{d}{dr} \\ 
-\frac{4\eta \left( \ell _{1}+\ell _{2}+\cdots +\ell _{d-1}\right) }{r}-\eta \frac{d-1+2\left( \mu _{1}+\mu _{2}+\mu _{3}+\cdots +\mu _{d}\right) }{r}+ \frac{2me^{2}}{\hbar ^{2}r}
\end{array}
\right\} \Xi \left( r\right) =0.
\end{equation}%
Here, we notice that the current differential equation of $\Xi \left( r\right) $ 
can be reduced to a confluent hypergeometric differential equation if the following substitution is made:
\begin{equation}
\zeta =2\eta r.
\end{equation}%
Once the elementary mathematical operations have been completed, we get the differential equation
\begin{equation}
\zeta \frac{d^{2}}{d\zeta ^{2}}\Xi \left( \zeta \right) +\left( B-\zeta
\right) \frac{d}{d\zeta }\Xi \left( \zeta \right) -A\Xi \left( \zeta \right)
=0,
\end{equation}%
which has solution in terms of the  confluent hypergeometric function  $\mathbf{F}\left(A, B; \zeta \right) $ with the arguments
\begin{eqnarray}
A&=&2\left( \ell _{1}+\ell _{2}+\cdots +\ell _{d-1}\right) +\left( \mu_{1}+\mu _{2}+\mu _{3}+\cdots +\mu _{d}\right) +\frac{d-1}{2}-\frac{me^{2}}{\eta \hbar ^{2}},  \\
B&=&4\left( \ell _{1}+\ell _{2}+\cdots +\ell _{d-1}\right) +2\left( \mu
_{1}+\mu _{2}+\mu _{3}+\cdots +\mu _{d}\right) +d-1.
\end{eqnarray}
Subsequently, the eigenfunctions can be represented 
\begin{equation}
U\left( r\right) =\mathcal{N}\text{ \ }r^{2\left( \ell _{1}+\ell _{2}+\cdots
+\ell _{d-1}\right) }e^{-\eta r}\mathbf{F}\left( A,B;2\eta r\right) ,
\end{equation}
where the normalization constant, denoted by $\mathcal{N}$, is yet to be determined. Skipping this, we now turn our attention to the energy spectrum. In consideration of the finite nature of solutions at infinity, we put forth the following quantum conditions
\begin{equation}
2\left( \ell _{1}+\ell _{2}+\cdots +\ell _{d-1}\right) +\left( \mu _{1}+\mu
_{2}+\mu _{3}+\cdots +\mu _{d}\right) +\frac{d-1}{2}-\frac{me^{2}}{\eta
\hbar ^{2}}=-n\text{, \ with\ \ }n\in \mathbb{N}.
\end{equation}%
From this, we obtain the energy spectrum as follows:
\begin{equation}
E_{n,\ell _{1},...,\ell _{d-1},d}=-\left( \frac{me^{4}}{2\hbar ^{2}}\right) 
\frac{1}{\left[ n+2\left( \ell _{1}+\ell _{2}+\cdots +\ell _{d-1}\right)
+\left( \mu _{1}+\mu _{2}+\mu _{3}+\cdots +\mu _{d}\right) +\frac{d-1}{2}%
\right] ^{2}}.
\end{equation}
We observe that for $d=3$ the resulting form of the obtained energy eigenvalue function becomes identical to that obtained in \cite{Bilel20221}. For a large-dimensional scenario, the energy eigenvalue function reduces to 
\begin{equation}
E_{n,\ell _{1},...,\ell _{d-1}}=-\left( \frac{2me^{4}}{\hbar ^{2}}\right) %
\left[ \frac{1}{d^{2}}-4\frac{n+2\left( \ell _{1}+\ell _{2}+\cdots +\ell
_{d-1}\right) +\left( \mu _{1}+\mu _{2}+\mu _{3}+\cdots +\mu _{d}\right) -1/2%
}{d^{3}}+...\right].
\end{equation}
This case suggests that the significance of the quantum numbers and Wigner parameters is diminished in higher dimensions. 

\begin{figure}[htb!]
\begin{minipage}[t]{0.5\textwidth}
        \centering
        \includegraphics[width=\textwidth]{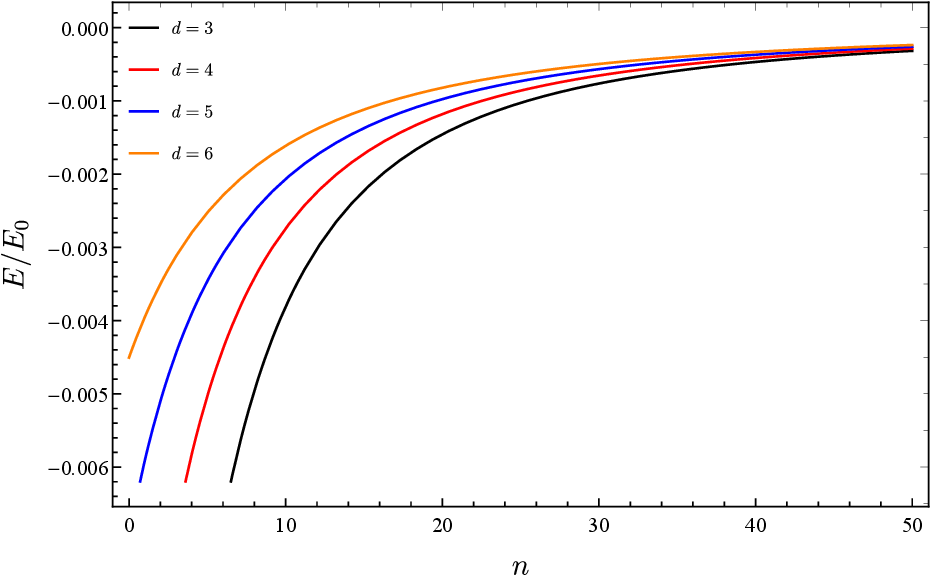}
        \label{fig:ba}
        \subcaption{$\mu_{i}=0.4$}
\end{minipage}
\begin{minipage}[t]{0.5\textwidth}
        \centering
        \includegraphics[width=\textwidth]{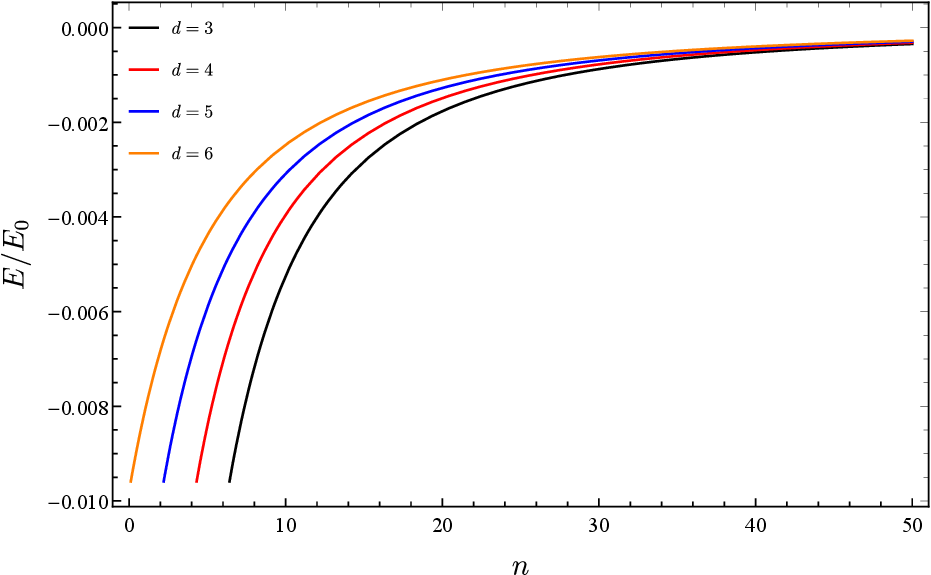}
       \label{fig:bb}
       \subcaption{$\mu_{i}=-0.4$}
   \end{minipage}
\caption{$\frac{E}{E_{0}}$ versus $n$ for various values of the spatial dimensions $d$.}
\label{fig:m}
\end{figure}

Finally, in Fig. \ref{fig:m} we illustrate the variation of the ratio of the energy value to the ground state energy according to the quantum number $n$ in various low dimensions. 

\newpage
We see that the effects arising from dimensions are significant at relatively small quantum numbers. 
\section{Conclusion}

The systems under investigation may yield disparate solutions contingent on the principle of parity. The conventional calculation techniques do not yield solutions that are directly dependent on parity. Nevertheless, research conducted in recent decades has demonstrated that utilizing the Dunkl operator in lieu of the conventional partial derivative operator facilitates the derivation of solutions that are dependent on parity. Furthermore, these solutions are influenced by additional free parameters, known as the Wigner deformation parameter. In this paper, we have solved the Dunkl-Schr\"odinger equation in any arbitrary dimension initially with the Cartesian coordinates. Subsequently, the d-dimensional polar coordinates were employed, demonstrating that the Dunkl-Schr\"odinger equation was separated into angular and radial parts, with the angular part equation solved in terms of Jacobi polynomials. The exact bound states solutions were then obtained for three systems: the isotropic harmonic oscillator, the pseudoharmonic oscillator, and the Coulomb potential. The results revealed that the energy eigenvalue functions are altered under the influence of Dunkl deformation in any arbitrary dimensions. Therefore, the deformation constant can be employed to achieve a precise alignment between experimental and theoretical outcomes.

\section*{Acknowledgments}
This work is supported by the Ministry of Higher Education and Scientific
Research, Algeria under the code: B00L02UN040120230003. B. C. L. is grateful to Excellence project PřF UHK 2211/2023-2024 for the financial support.

\end{document}